\documentclass[preprintnumbers,amsmath,amssymb,floatfix,11pt,prd,onecolumn,
superscriptaddress,nofootinbib]{revtex4}
\usepackage{latexsym,float}
\usepackage{epsfig}
\usepackage{epstopdf}
\usepackage{caption}
\usepackage{subfig,color}
\usepackage{amssymb}
\usepackage{amsmath, hyperref}
\usepackage{bm}

\begin{document}

\title{On minimal energy states of chiral MHD turbulence}

\author{Petar Pavlovic}\email{petar.pavlovic@desy.de}

\author{and G\"unter Sigl} \email{guenter.sigl@desy.de}

\affiliation{II. Institut f\"ur Theoretische Physik, University of Hamburg\\
Luruper Chaussee, 149, 22761 Hamburg, Germany}

\begin{abstract}
We study the evolution of magnetohydrodynamic turbulence taking into account the chiral anomaly effect. This chiral magnetohydrodynamic
description of the plasma is expected to be relevant for temperatures comparable to the electroweak scale, and therefore for the evolution of magnetic fields in the early Universe 
and young neutron stars. We focus on the case of freely decaying chiral magnetohydrodynamic turbulence and
discuss the dissipation of ideal MHD invariants. Using the variational approach we discuss the minimum energy configurations of magnetic
field and velocity. As in the case of the standard magnetohydrodynamic turbulence, we find that the natural relaxation state is given
by a force-free field, $\nabla \times \mathbf{B} \propto \mathbf{B}$. However, the precise form of this configuration is now determined by parameters describing the chiral anomaly effect, leading to some
important differences when compared to the non-chiral turbulence. Using this result we argue that the evolution of velocity and magnetic field will
tend to effectively decouple during the transition to this minimal energy state. 
\end{abstract}  

\maketitle

\section{Introduction}
\label{sec:intro}
The existence of turbulent magnetic fields has been observed on almost all scales of the observable Universe -- including solar processes and solar wind,
accretion disks, and galaxy clusters \cite{magso, gold, armstrong, scalo, Vogt, R.Durrer}. Magnetic fields are also expected 
to be produced in the context of the early Universe, where they can influence various important
processes such as primoridal nucleosynthesis \cite{greenstein,cheng,grasso}, creation of gravitational waves 
\cite{grigol, alberto, chiara, gerold} and cosmic microwave background (CMB) spectrum anisotropy patterns
\cite{zeldovich,grishchuk,gasperini}. One of the important problems in this context is the evolution of
turbulent magnetic fields in the cosmological setting \cite{wag,bran1,ol,son}. Description of turbulent magnetized plasma is usually given in terms of magnetohydrodynamic (MHD) 
approximation --a theoretical model which uses the set consisting of Maxwell, Navier-Stokes and continuity equation, while assuming the global neutrality
of plasma and fluid approximation. \\ \\
However, even on macroscopic scales, the standard MHD description of turbulence should be modified for the temperatures comparable
to the electroweak scale, due to the quantum effect of chiral anomaly -- which leads to the coupling between 
velocity, magnetic fields and the particle content of the theory. In the case of massless charged particles placed in a (hyper)magnetic field -- for instance leptons before the electroweak transition -- the chiral
magnetic effect leads to the creation of additional effective current, in the direction of the magnetic field and proportional to the difference between the number of left and right-handed
chiral particles \cite{Vilenkin, nielsen, fukushima, Kharzeev, sadofyev}:
Even if there are no particle processes which would cause transitions between the chiral states, the difference of respective
chemical potentials of left-handed and right-handed particles, $\mu_{5}=(\mu_{L}-\mu_{R})/2$, will not be conserved, but will change according to \cite{Kharzeev}
\begin{equation}
\frac{d \mu_{5}}{d t}=\frac{1}{T^{2}} \frac{3e^2}{4 \pi^2}\frac{d h}{d t}, 
\label{chempo}
\end{equation}
where $T$ stands for temperature of the system, and $h$ is the magnetic helicity density defined as $h=V^{-1} \int \mathbf{A} \cdot \mathbf{B} d^{3}r$, where $\mathbf{A}$ is the 
vector potential and $\mathbf{B}$ is the magnetic field. When the processes
which flip the particle chirality are also present, the rate of such processes can be perturbatively added to the equation \ref{chempo}, which will then act as an additional source
of non-conservation of $\mu_{5}$. The change of the chiral chemical potential in the magnetic field, given by \eqref{chempo}, then corresponds to the effective chiral current, which should also be included in the MHD equations
 \begin{equation}
\mathbf{j_{5}}=-\frac{e^{2}}{2 \pi^{2}}\mu_{5} \mathbf{B}. 
\label{current}
\end{equation}
Note that this effective current, created by the chiral magnetic effect, is a vector current and is thus naturally added to the standard electric current. 
The chiral anomaly effect was recently explored in heavy ion collisions \cite{khar}, early universe \cite{Boyarsky, hiro, pa, gorbar} and neutron stars \cite{Ohnishi:2014uea, emass, Leite, kaminski},
as well as in the context of magnetogenesis and leptogenesis \cite{sha, gi, dvo, sem, kamada, kamada2}. \\ \\
This modification of MHD equations, that comes as a result of the chiral anomaly effect, naturally opens the question of extending the study of MHD turbulence to the case of 
chiral MHD turbulence. Such generalization leads to the complex interplay between the velocity field and magnetic field evolution -- that are now influenced by turbulence and the evolution of chiral potential, as well as the particle physics involved -- which determines the rates of flipping processes and thus the evolution
of the chiral asymmetry. In fact, it seems very reasonable to assume the existence of velocity field in the early Universe, coming from the phase transitions or density perturbations 
 \cite{kam,kos,tina, wag}, which will typically be associated with large Reynolds numbers, $Re=Lv/\nu$ (with the characteristic length scale $L$, characteristic velocity $v$ and a kinematic viscosity $\nu$), typically leading to turbulence. Therefore, the description of chiral MHD turbulence should be considered as a necessary part of understanding the physical
 processes around the electroweak scale, both in the early Universe and astrophysics \cite{pa, Leite, novi}. However, the issue of of chiral MHD turbulence cannot be addressed in a simple manner -- due to the mathematical 
 complexity of the chiral electrodynamic equations as well as the non-linearity of the Navier-Stokes equation, which makes even the hydrodynamical turbulence to be one of the unsolved
 problems of classical physics. It is therefore necessary that the study of chiral MHD turbulence is based on the simplified theoretical models and numerical simulations. This
 interesting and difficult problem was addressed only recently. Scaling laws of the chiral MHD turbulence were proposed in \cite{yam}, based on the scaling symmetries. 
 Chiral turbulence was further analysed in \cite{novi}, where it was discussed how the chiral magnetic effect leads to the creation of helical magnetic fields and changes the evolution
 of magnetic energy and correlation length, while supporting the inverse cascade. In \cite{Sem} chiral turbulence around the electroweak transition was discussed, using the approximation
 for the velocity field based on the drag time parameter. Theoretical considerations of dynamos in chiral magnetohydrodynamics can be found in \cite{roga}, while the numerical
 study of chiral MHD turbulence in the early universe was recently reported in \cite{brandy}, \cite{jennifer} and \cite{jennifer2}. \\ \\
 As noted before, the complexity of turbulence makes the focus on special states and regimes necessary in analytical studies. One of such settings is the 
 force-free MHD configuration which can be realized as a minimum-energy state of the MHD turbulence. This approach was in the focus of considerable interest and numerous works
 in the field of plasma physics, especially on the subject of toroidally confined pinch plasmas \cite{plazma1, plazma2, plazma3, plazma4}, and was also experimentally
 investigated \cite{plazma5}. The same approach was followed in theoretical and experimental studies related to MHD of solar corona \cite{plazma5, plazma6}. \\ \\
 The aim of this paper will be to generalize the discussion on the minimal energy MHD configurations to the case of chiral anomaly turbulence. We will be particularly interested
 in analyzing the changes which the chiral anomaly effect creates in the minimal energy states with respect to standard MHD turbulence. 
 \\ \\
 This work is organized in the following manner: in \S\ref{sec:ModMHD} the MHD equations in the presence of the chiral anomaly are reviewed and introduced and the dissipation of important MHD quantities is discussed; in \S\ref{sec:min} we discuss the minimal energy states of chiral MHD turbulence; and we conclude in \S\ref{sec:conc} .

\section{Chiral MHD equations and ideal invariants} \label{sec:ModMHD}
Introducing the coefficient 
$c_{5}=e^{2}/4 \pi^2$   
we can write the modified chiral MHD equations in Lorentz-Heaviside units in the following form \cite{novi}
\begin{equation} 
 \nabla \times \textbf{B} = \sigma \left( \textbf{E} - 2 \frac{c_{5}}{\sigma}   \mu_{5} \textbf{B} + \textbf{v} \times \textbf{B}\right),
\label{Maxwell}
\end{equation}
\begin{equation}
 \frac{\partial \textbf{B}}{\partial t}= - \nabla \times \textbf{E},
 \label{rot}
\end{equation} 
 \begin{equation}
 \rho\left[\frac{\partial \textbf{v}}{\partial t} + (\textbf{v} \cdot \nabla) \textbf{v} - \nu \nabla^{2}\textbf{v}\right]= - \nabla p + \left[\sigma \textbf{E}\times \textbf{B} + (\textbf{v} \times \textbf{B}) \times \textbf{B}\right],
\label{vel}
\end{equation}
\begin{equation}
\frac{\partial \rho}{\partial t}+ \nabla(\rho \cdot \textbf{v})=0,
\label{c}
\end{equation}
\begin{equation}
\frac{d \mu_{5}}{d t}=\frac{3}{T^{2}} c_{5} \frac{d h}{d t} - \Gamma_{f} \mu_{5}, 
\label{mutot}
\end{equation}
where  $\rho$ is the matter density, $\sigma$ is the conductivity, $\mathbf{v}$ is the velocity, $\nu$ is the kinematic viscosity $\Gamma_{f}$ is the total chirality flipping rate, and we assume that there are no additional source terms active,
which would generate a chiral asymmetry $\mu_{5}$ in the considered system. Such assumption necessary follows if one wants to discuss 
minimal energy states, since the presence of a source term would act as an additional exterior source of energy transferred into the system, 
and it would thus oppose the relaxation process. $\textbf{E}$ and $\textbf{B}$ denote the electric and magnetic field, respectively. 
Furthermore, the global neutrality of plasma is assumed, i.e. $\nabla \cdot \textbf{J}=0$ and $\nabla \cdot\textbf{E}=0$, and the displacement current is neglected.  
We focus on the case of the incompressible fluid (for a discussion of this assumption see \cite{novi}). As it is well known, the MHD equations will
have the same form on curved spacetime as long as time is replaced by conformal time and all the physical quantities are scaled with the conformal factor \cite{Banarjee}, $a(t)$, so they
can be simply applied to the case of the early Universe. \\ \\
The chiral MHD equations of this form were commonly used in the study of the chiral magnetic effect applied to the problems of the 
early universe and young neutron stars (see the references in the Introduction). They assume that the global motion of plasma can be 
treated as completely non-relativistic, so that the chiral modifications can be approximately described as an addition of the effective
current associated with the chiral magnetic effect to the equations of the non-relativistic macroscopic MHD. The realistic systems of interest
will furthermore typically be characterized by the condition $\mu_{5} /T \ll 1$ \cite{pa, novi}. Although one should in general expect further corrections and additional terms arising from various chiral effects, under these conditions they will all typically be negligible with respect to the terms included in the equations \ref{Maxwell} -- \ref{mutot}. As a consequence of this relative simplicity of the equations it is possible to analytically discuss the complicated issue of relaxation processes related to chiral MHD turbulence as will be done in this work. Let us now 
discuss different potential corrections that could modify the considered set of equations \ref{Maxwell} -- \ref{mutot}. First of all, one could consider 
the contribution of the effective anomalous current to the energy-momentum tensor, which would then modify the Navier-Stokes equation. 
However, considering the correction of the anomaly contribution to the energy current -- which is proportional to $\mu_{5}^2 + T^{2}$ \cite{land} - it can be 
clearly seen that the anomaly contribution is negligible in the regimes of our interest where $\mu_{5}/T \ll 1$. 
On the other hand, there is an additional chiral effect, related to the induction of the effective current along the direction of vorticity, and
is thus known as the chiral vortical effect -- which was also extensively studied in the literature \cite{ A., black, Surowka, Megias}. The interesting 
consequence of this effect is the possibility to transfer the energy between the kinetic sector, associated with the movement of the fluid, 
and the chiral sector. However, while the chiral magnetic effect is proportional to $\mu_{5}$, the vortical effect is actually proportional 
to $\mu_{L}^{2} - \mu_{R}^{2}$ \cite{wang}. Therefore, in the conditions on which we restrict our attention here, where the respective chemical potentials are significantly smaller
than the temperature of the system, the chiral vortical effect
will typically be negligible with respect to the chiral magnetic effect. In further studies on the chiral turbulence it could be interesting 
to try to approach other regimes apart from the one where the chiral potential is much smaller than the temperature, but it is questionable 
how much it could be treated analytically in general. Let us finally note that the same considerations of non-relativistic velocities and the chiral potential being much smaller than the temperature, also imply that the contribution of the chiral magnetic and chiral vortical effect to the 0th components of the four current, discussed in \cite{Surowka}, will not be of interest in the current discussion.  \\ \\
The system of equations \eqref{Maxwell}-\eqref{mutot} will lead to the MHD turbulence for high enough values of kinetic and magnetic Reynolds number, $Re\gg 1$ and $Re_{B}\gg 1$, which are defined respectively as
$Re=L v/\nu$ and $Re_{B}\equiv \ 4 \pi Lv \sigma$. One can also easily define various quantities of physical interest for discussing the evolution of chiral turbulence, which will
be used in the remainder of this work. The energy of magnetized fluid is described by the magnetic energy density
\begin{equation} 
\rho_{m}=\frac{1}{2V} \int d^{3} r \mathbf{B}^{2}(\mathbf{r},t),
\label{magnetic}
\end{equation}
and the kinetic energy density
\begin{equation}
\rho_{K}=\frac{1}{2V}\int d^{3} r \rho  v^{2} .
\label{kinetic}
\end{equation}
In the case of MHD turbulence it is also necessary to take into account the contribution coming from  the chiral chemical potential, $\rho_{5}$. Under the condition of applicability of chiral
MHD, $\mu_{5} \ll e^{2}T$ \cite{yam}, and using the definition of chemical potential this contribution can be written as
 \begin{equation} 
\rho_{5}=\frac{\mu_{5}^{2}T^{2}}{6}
\label{chirale}
\end{equation}
Apart from the total energy, $\rho_{tot}=\rho_{m} + \rho_{K} + \rho_{5}$ the chiral fluid is also characterized by the so-called ideal invariants, the quantities which are conserved
in the case of the ideal MHD, i.e when $\sigma \rightarrow \infty$, $\nu \rightarrow 0 $. Magnetic helicity, already defined before, describing the linking and twisting of magnetic field lines, represents
such a quantity, together with the cross-helicity density:
\begin{equation}
h_{c}= \frac{1}{V}\int d^{3} r \mathbf{v} \cdot \mathbf{B}
\end{equation}
Although these quantities are conserved only in the ideal case, they also play an important role in the resistive regime, influencing the organization of turbulent structures,
time evolution of turbulence and the existence of inverse cascades \cite{novi}. These effects are closely associated with the fact that decay rates of total energy on the one hand, and
the magnetic and cross helicity on the other hand, are different -- so that the helicities can be effectively considered as approximately invariant for high conductivities. 
This decay of ideal MHD quantities at different rates, called selective decay, is a well studied phenomenon in the context of MHD turbulence \cite{poq, bis, math} and we argue that it should also
be applied to the problem of chiral MHD turbulence.\\ \\
With this objective in mind, we first study the dissipation of ideally conserved quantities in the chiral MHD case. In order to obtain the expression for dissipation of magnetic and
kinetic energy we multiply the Navier-Stokes equation \eqref{vel} with $\mathbf{v}$ and Eq. \eqref{rot} with $\mathbf{B}$, adding them using definitions \ref{magnetic} and 
\ref{kinetic}. Then, expressing the pressure divergence as a function of other quantities in the incompressible limit, $\nabla^2 p=-\nabla \cdot \left[ (\mathbf{v} \cdot \nabla)\mathbf{v}- (\mathbf{B} \cdot \nabla)\mathbf{B} \right] $,
and ignoring the surface terms as usual we obtain
\begin{equation}
\frac{d}{dt}(\rho_{m}+\rho_{K})= - \frac{1}{V} \int\left[\frac{(\nabla \times \mathbf{B})^{2}}{\sigma} + \nu \cdot \omega^{2}+ \frac{2c_{5}}{\sigma} \mu_{5} \mathbf{B} \cdot (\nabla \times \mathbf{B} +\mathbf{J}_{5})\right] d^{3}r,  
\label{maghyen}
\end{equation}
where $\mathbf{\omega}$ is the vorticity defined as $\boldsymbol{\omega}= \nabla \times \mathbf{v}$, and
we introduced the effective chiral current $\mathbf{J}_{5}=2 c_{5} \mu_{5}  \mathbf{B}$. Here we have also used the fact that 
$\mathbf{B} \cdot \nabla^{2} \mathbf{B}=- (\nabla \times \mathbf{B})^{2}$ and $\mathbf{v} \cdot \nabla^{2} \mathbf{v}=-\omega^{2}$ in the incompressible limit, and expressed the quantities in the units of
Alfven time, so that $\rho=1$. 
We see that, as in the standard MHD case, the dissipation of magnetohydrodynamic energy is determined by the resistivity and viscosity of the fluid. On the other hand, the chiral anomaly
contribution can, depending on the sign of $\mu_{5}$ and relative orientation of the magnetic field and the effective current, lead to the growth of magnetic energy (when the energy stored in the
chiral asymmetry chemical potential gets transferred to magnetic energy via the chiral anomaly effect), or its further dissipation (when the magnetic energy transforms to energy stored in the
chiral asymmetry). All of these contributions vanish in the limit of infinite conductivity and zero viscosity, where the energy is a conserved quantity.\\ \\ 
Using the definition of magnetic helicity and equations \eqref{Maxwell} and \eqref{rot} it can be shown that the change of magnetic helicity is given by
\begin{equation}
\frac{dh}{dt}=-\frac{2}{V \sigma} \int d^3{r}
(\mathbf{\nabla \times \mathbf{B}}+ \mathbf{J}_{5}) \cdot \mathbf{B},
\label{helchange}
\end{equation}
so that it consists of two contributions -- the dissipation due to the resistivity of the medium, and the chiral anomaly contribution, both of which
vanish in the limit of infinite conductivity. As it was the case with the magnetic energy, the chiral effect leads to the enhanced violation of helicity 
in the limit of finite resistance and supporting its growth. We can now use \eqref{chirale}, together with \eqref{chempo} to determine the time change of total energy, $\rho_{tot}$. Using the equation
\eqref{helchange} we note that the change of energy stored in the chiral assymetry potential exactly cancels the anomaly created change in \eqref{maghyen}, as required by the 
conservation of energy, so that
\begin{equation}
\frac{d \rho_{tot}}{dt} =  - \frac{1}{V} \int \left[\frac{(\nabla \times \mathbf{B})^{2}}{\sigma} + \nu \cdot \omega^{2}\right] d^{3}r -\frac{T^{2}}{3} \mu_{5}^{2} \Gamma_{f} .
\end{equation}
The change of the total energy is guided by three dissipation processes, and it is therefore strictly negative. Apart from the dissipation due to the resistivity and viscosity, in the chiral case
the total energy also decreases due to the chirality flipping processes, which decrease the chiral assymetry, and therefore the energy initially stored in it. When the rates of flipping 
processes are negligible, the total change of energy is equal to the change of energy in the non-chiral MHD.\\ \\
Finally, we can also calculate the time change of cross-helicity, $h_{c}$ to be
\begin{equation}
\frac{dh_{c}}{dt}= -\frac{1}{V} \int\left[(\nu + \frac{1}{\sigma}) \cdot(\nabla \times \mathbf{B}) \cdot \boldsymbol{\omega}  + \frac{2c_{5}}{\sigma} \mu_{5} \cdot (\nabla \times \mathbf{B} +\mathbf{J}_{5})\cdot \mathbf{v}\right]d^{3}r ,
\end{equation}
and we see that, when compared to the cross-helicity evolution in the standard MHD, here it also gets modified due to the chiral effect. The interesting consequence of this relation is that,
similar to the creation of magnetic helicity due to the chiral effect \cite{pa, novi}, the chiral anomaly can also lead to the creation of cross-helicity, even if the chiral
MHD turbulence initially has a vanishing cross-helicity. This is consistent with the results already discussed in \cite{yam}.   \\ \\
Since the dissipation of ideal MHD invariants is related to the quantities like $\nabla \times \mathbf{B}$ and $\boldsymbol{\omega}$, which include the spatial variations of the magnetic field and velocity,
it is obvious that dissipation of total energy will not have the same rate as the dissipation of magnetic helicity and cross-helicity. Apart from different order of such quantities
appearing in the dissipation equation for the energy and magnetic helicity, both helicities can increase in time due to the anomaly contribution, while the total energy needs to strictly
decrease with time. We can therefore conclude that during the relaxation of the chiral MHD system towards the minimal energy configuration the dissipation of total energy will in principle
be faster than the dissipation of magnetic and cross-helicity, and this difference can increase due to the chiral anomaly effect. Thus, one of the interesting consequences of
chiral anomaly in the context of turbulence is that it can influence and enhance the selective decay process. 

\section{Minimal energy configurations}\label{sec:min}
Following the conclusions from the previous section, we conclude that there will two possible relaxation processes - the relaxation towards a minimal energy state characterized by the
constant magnetic helicity or the minimal energy state characterized by the constant cross-helicity. The first type of transition 
means that at the later stages of chiral turbulence the chiral anomaly contribution will become negligible, since the change in the chiral asymmetry is directly related to the change in helicity, and will thus be 
negligible. This type of chiral MHD evolution, characterized by a phase where the magnetic helicity saturates after the anomaly dominated magnetic field growth, was also recently reported in numerical studies of chiral MHD turbulence \cite{jennifer, brandy}. We will first consider this type of selective decay process. 
If we demand that the minimal energy configuration is established with respect to the constraint of constant helicity, we need to perform the variational principle of the following expression
\begin{equation}
J= \frac{1}{V}\int \left[\frac{1}{2}  (v^{2} +(\nabla \times \mathbf{A})^{2}))  +  \frac{\mu_{5}^{2}T^{2}}{6}\right] d^{3} r -  \frac{\alpha}{2V} \int \mathbf{A} \cdot (\nabla \times \mathbf{A}) d^{3}r,
\label{variational}
\end{equation}
where $\mathbf{A}$ is the vector potential, and $\alpha$ is the Lagrangian multiplier. Varying \eqref{variational} with respect to velocity while requiring 
the extremal configuration
simply yields $\mathbf{v}=0$, which signifies that velocity will typically dissipate approaching the zero velocity case, as its minimal energy configuration. We next perform the variation with respect 
to the vector potential. Ignoring the rates of flipping reactions, 
$\Gamma_{f}=0$, and using \ref{chempo} assuming a constant temperature $T$ \footnote{We note that this assumption does not present any limitation for the application
of this approach to the case of the early Universe -- where temperature clearly needs to change, since introducing the description in terms of conformal quantities, temperature does not appear explicitly in  \eqref{variational}, and the treatment is mathematically the same, after replacing the ordinary time with conformal time and variables with their conformal counterparts  (for instance see \cite{pa}) }, while varying \eqref{variational} with respect to vector potential, using the identity 
$\mathbf{A} \cdot (\nabla \times \delta \mathbf{A})=\delta \mathbf{A} \cdot(\nabla \times \mathbf{A}) - \nabla \cdot (\mathbf{A} \times \delta \mathbf{A})$, we get
\begin{equation}
\nabla \times \mathbf{B} + \left[2c_{5}(\frac{3c_{5}}{T^{2}}h +C)- \alpha\right] \mathbf{B}=0, 
\label{varhel}
\end{equation}
where
\begin{equation}
 C=\mu_{5}^{in} - \frac{3c_{5}}{T^{2}} h^{in},
 \label{constant}
\end{equation}
and $\mu_{5}^{in}$, $h^{in}$ are the initial values of chiral asymmetry potential and magnetic helicity long before the relaxation stars. Note that
the approximately constant helicity in the relaxation phase will in general not be the same as the helicity characterizing the 
beginning of chiral turbulence ($h^{in}$), since chiral turbulence will, as discussed, typically have other phases before the relaxation phase
where the value of magnetic helicity can change significantly. This entering into the relaxation phase with approximately constant 
helicity can be expected at late times at which $\mu_{5}$ is already significantly depleted due to the transfer of energy to the magnetic 
sector, such that the only significant changes in magnetic helicity will come from the effects of fine resistivity (for particular 
realizations of such chiral MHD turbulence scenarios see the results of simulations in \cite{brandy, jennifer, jennifer2}). 
The considered relaxation state was achieved with respect to the constraint of
constant magnetic helicity, so the helicity value appearing in \eqref{varhel} is the value towards which helicity saturates during this selective decay. Thus, during this relaxation process the system approaches
a force-free field configuration, which is determined by the parameters describing the chiral anomaly effect. Physically speaking, the only case in which the evolution of chiral MHD turbulence and the standard MHD turbulence coincide
is for $\mu_{5}^{in}=0$ and $h^{in}=0$ -- and this can be satisfied only for the non-helical turbulence. However, one can make a transition to the classical MHD description of turbulence in the general case 
if the quantum anomaly effect, relating the change of magnetic helicity to the change of chiral chemical potential, is ignored. This classical limit is formally achieved if the characteristic coupling 
coefficient is taken to be vanishing, $c_{5}=0$, and it can be seen that then all the equations reduce to the ones characteristic for the standard MHD. This will physically correspond 
to the case when the effective chiral curent is much smaller than the total current, $|\mathbf{J}| \ll |\nabla \times \mathbf{B}| \approx |\mathbf{B}/l|$ and thus $c_{5} \ll 1/(\mu_{5} l)$, where
$l$ is a characteristic scale over which the magnetic field changes. In the case of standard MHD turbulence, i.e. for $c_{5}=0$, we obtain the condition 
$\nabla \times \mathbf{B} - \alpha \mathbf{B}=0$. In this case the curl of the magnetic field can only vanish if the field itself vanishes for $\alpha \neq 0$, while for the chiral case
the configuration corresponding to a vanishing curl of the magnetic field is also possible if $(2c_{5})(3c_{5}/T^{2}h +C)= \alpha$. In the non-chiral case the orientation of curl of $\mathbf{B}$ can be either parallel (for $\alpha >0$)
or anti-parallel ($\alpha <0$) for all field configurations, while in the chiral case both orientations are possible for a given value of $\alpha$. It is possible to express the constant $\alpha$ in terms of the magnetic energy characterizing the minimal configuration, $\rho_{m}^{min}$, magnetic helicity, 
and the remaining chiral parameters, taking a dot product of \eqref{varhel} with $\mathbf{A}$, integrating both sides over volume, and using definitions of magnetic and helicity energy densities:
\begin{equation}
\alpha = \frac{2\rho_{min}}{h}- 2c_{5}(\frac{3c_{5}}{T^{2}}h +C)
\end{equation}
\\
Let us now briefly discuss the case where the system already evolved sufficiently close to this minimal-energy configuration. Approaching the discussed state, characterized by 
$\mathbf{v}=0$, the velocity will dissipate and the second-order quantities in velocity appearing in the Navier-Stokes equation can be ignored, while the field configuration will be close to 
a force free configuration. Then the evolution of velocity will be approximately given by
\begin{equation}
\frac{\partial \textbf{v}}{ \partial t}  - \nu \nabla^{2}\textbf{v}= 0,
 \label{san}
\end{equation}
and thus decouples from the evolution of magnetic field in this approximation. 
We can now use the Fourier decomposition
\begin{equation}
\mathbf{v}(\mathbf{r},t)=\int \frac{d^{3} q}{(2 \pi)^{3}}e^{i\mathbf{q} \cdot{\mathbf{r}}}\mathbf{v}(\mathbf{q},t)  
\end{equation}
to the solutions for the velocity modes
\begin{equation}
\mathbf{v}_{\mathbf{q}}(t)=\mathbf{v_{q}}(0) e^{-\nu q^2 t},
\label{veli}
\end{equation}
were $\mathbf{v_{q}}(0)$ is an integration constants. Using the Fourier decomposition of the magnetic field,
\begin{equation}
\mathbf{B}(\mathbf{r},t)=\int \frac{d^{3} k}{(2 \pi)^{3}}e^{i\mathbf{k} \cdot{\mathbf{r}}}\mathbf{B}(\mathbf{k},t) 
\end{equation}
in equations \eqref{Maxwell} and \eqref{rot} we obtain
 \begin{equation}
 \partial_t \mathbf{B_k}=-\frac{k^2}{\sigma}  \mathbf{B}_\mathbf{k} - \frac{2}{\sigma} \mu_5 (i\mathbf{k}\times \mathbf{B}_\mathbf{k})  
 +\mathbf{I}  \, ,
 \label{magfi}
\end{equation}
where the evolution of chiral magnetic field is modified with respect to the one studied in \cite{Boyarsky, pa} by the addition of turbulent term
\begin{equation}
 \mathbf{I}=\frac{i}{(2\pi)^{3/2}}\mathbf{k}\times\int d^3q (\mathbf{v}_\mathbf{k-q}\times \mathbf{B}_\mathbf{q}) . 
\end{equation}
Taking the scalar product of the equation \eqref{magfi} with $B^{i}_{\mathbf{k}}$, and then taking the ensemble averages, we can derive the equations for the time change
of magnetic energy and helicity modes, as in \cite{Boyarsky, pa, novi}. In the current regime, of the MHD turbulence approaching its minimal energy state dictated by
the conservation of helicity, using \eqref{veli} we also obtain the turbulent contribution to the time change of the energy density of the following form
\begin{equation}
\frac{i}{(2\pi)^{3/2}}\int d^3q e^{-\nu q^2 t} [k_{a}v^{i}_{\mathbf{k}-\mathbf{q}}(0)<B^{a}_{\mathbf{q}}B^{i}_{\mathbf{k}}> - 
k_{a}v^{a}_{\mathbf{k}-\mathbf{q}}(0)<B^{i}_{\mathbf{q}}B^{i}_{\mathbf{k}}>]
\end{equation}
Focusing our attention on the evolution of statistically homogeneous and isotropic magnetic fields we use the condition 
\begin{equation}
<B_{i}(\mathbf{k},t) B_{j}(\mathbf{q},t)>=\frac{(2 \pi)^3}{2} \delta(\mathbf{k}+\mathbf{q})\left[(\delta_{ij}-\frac{k_{i} k_{j}}{k^2})S(k,t) + i \epsilon_{ijk}\frac{k_{k}}{k}A(k,t)\right], 
\label{correlator}
\end{equation}
where $S$ and $A$ denote the symmetric and antisymmetric parts of the correlator. Applying this condition and using the assumption of incompressibility, we see that 
 the contribution of the last term vanishes, and we are left only with the non-velocity chiral MHD equations:
 \begin{equation} 
\partial_t  \rho_k= - \frac{2 k^2}{\sigma} \rho_k - 4 \frac{c_{5}}{\sigma} \mu_5 k^2 h_k \, ,
\label{rhok}
\end{equation}  
\begin{equation} 
 \partial_t  h_k= - \frac{2 k^2}{\sigma} h_k - 16 \frac{c_{5}}{\sigma} \mu_5 \rho_k  \, .
 \label{hok}
 \end{equation}
We see that the evolution of chiral magnetic field and turbulence will tend to effectively decouple during the transition towards a minimal energy state under the constraint of constant magnetic helicity. \\ \\
We will now consider the second type of relaxation process, where the transition towards minimal energy state is determined by the constraint of constant cross-helicity. Which of the two possible relaxations will happen in a considered system seems to be determined by
the specific configuration of initial conditions. If the magnetic field is strongly helical it will prefer a transition towards a minimal energy state under the constraint of constant magnetic helicity, whereas if there is a significant initial alignment between the 
velocity and magnetic field the second case of constant cross-helicity will be preferred \cite{Biskamp,tel}. In the case of approximately constant cross-helicity we need to vary the following expression
\begin{equation}
J{'}= \frac{1}{V}\int \left[\frac{1}{2}  (v^{2} +(\nabla \times \mathbf{A})^{2}))  +  \frac{\mu_{5}^{2}T^{2}}{6}\right] d^{3} r -  \frac{\alpha'}{V} \int \mathbf{v} \cdot (\nabla \times \mathbf{A}) d^{3}r.
\label{cross}
\end{equation}
Varying \ref{cross} with respect to velocity while demanding $\delta J'=0$ we get
\begin{equation}
\mathbf{v}= \alpha' \mathbf{B},  
\label{jedn1}
\end{equation}
and varying with respect to the vector potential leads to
\begin{equation}
\nabla \times \mathbf{B} + 2c_{5}(\frac{3c_{5}}{T^{2}}h + C)\mathbf{B}= \alpha' (\nabla \times \mathbf{v}).  
\label{jedn2}
\end{equation}
In the special case when $\alpha'=\pm 1$, under the condition $\mathbf{B}\neq 0$ equations \ref{jedn1} - \ref{jedn2} have a solution only if the chiral effect is not present, formally $c_{5}=0$, and thus this state is not achievable in the chiral MHD turbulence. In the opposite case, when $|\alpha'|\neq 1$, the magnetic field obviously satisfies 
\begin{equation}
\nabla \times \mathbf{B}= -2 \frac{c_{5}(\frac{3c_{5}}{T^{2}}h + C)\mathbf{B}}{1-\alpha'^{2}} 
\end{equation}
If we take a dot product of this equation with $\mathbf{A}$ and integrate over volume, we can express the coefficient $\alpha'$ in terms of the magnetic energy density characterizing the minimal energy
configuration, $\rho_{min}$ and helicity at saturation as
\begin{equation}
\alpha'=\pm \sqrt{1+\frac{2c_{5}}{\rho_{min}} {(\frac{3c_{5}}{T^{2}}h + C)h}}. 
\end{equation}
In order to see the physical implications of this result better, we can write the Navier-Stokes equation \ref{vel} in terms of the Els\"{a}sser variables, $\mathbf{z}^{\pm}=\mathbf{v} \pm \mathbf{B}$
\begin{equation}
\frac{\partial z^{\pm}}{\partial t}+ z^{\mp}\cdot \nabla z^{\pm}= - \nabla p + \frac{1}{2}(\nu + \eta) \nabla^{2} z^{\pm} + \frac{1}{2}(\nu - \eta) \nabla^{2} z^{\mp} .
\end{equation}
In the non-chiral MHD case, corresponding to $\mathbf{v}= \pm \mathbf{B}$, the non-linear interaction term is necessarily zero, since it couples the Els\"{a}sser variables of different
signs -- in this case the evolution of turbulence is therefore dictated only by the dissipation processes. On the other hand, in the chiral MHD turbulence case we see that the interaction term will in general 
still be present. We conclude that the introduction of chiral anomaly effect leads to presence of non-linear interactions even in the minimal energy configuration for constant 
cross-helicity.

\section{Discussion and conclusion}\label{sec:conc}
In this work we have analyticaly studied the problem of relaxation towards a minimal energy state in the context of chiral MHD turbulence. We have first discussed the dissipation of
ideally conserved MHD quantities and showed that the chiral anomaly effects can further enlarge the difference between the decay of total energy and the magnetic/cross helicity.
While discussing the dissipation processes in the framework of chiral MHD turbulence, it was demonstrated how the chiral anomaly leads to a transfer of energy between the chiral
asymmetry chemical potential, $\mu_{5}$, and the magnetic field, such that the total energy related to this process remains conserved, while the dissipation of the total energy in
the system happens due to the viscous, resistive and chirality flipping processes. Using the fact that the magnetic and cross helicity in general dissipate slower in comparison to
the magnetic energy, we have discussed minimal energy configurations using a variational approach. There are two such relaxation states possible -- with respect to approximately
conserved magnetic helicity and with respect to approximately conserved cross-helicity. The influence of the flipping rates was neglected in this analysis, since, despite their
potentially complicated functional dependence, their effect would simply be to damp the chiral asymmetry and to subsequently cause the transition towards minimal energy states of the standard MHD.
The actual relaxation process happening in a system seems to be determined by the initial conditions, which
then favor one or the other possibility. While this minimal energy state, as in the standard MHD case, is given by a force-free field, its precise form includes
the corrections coming from the anomaly terms. Furthermore, in the case of a minimal energy state obtained with respect to approximately conserved cross-helicity, the velocity and magnetic
field will no longer be aligned such that $\mathbf{v}= \pm \mathbf{B}$, but will be related via  $\mathbf{v}= \pm \alpha' \mathbf{B}$, where $\alpha$ depends on the coefficients characterizing
the chiral anomaly, reducing to $\alpha'=1$ only if the anomaly effects are not active. This result leads to the presence of non-minimal interactions between velocity and magnetic field in such a state, which
are caused by the anomalous effects. The relaxation towards a minimal-energy state with respect to constant magnetic helicity resembles the later phase in the evolution of chiral MHD turbulence
as reported in recent numerical studies. These results suggest that the evolution of chiral MHD turbulence, characterized by the initial helicity and sufficiently strong asymmetry
between the left and right-handed charged particles, after the initial anomaly dominated phase subsequently typically enters into the phase of approximately constant helicity, approaching
the minimal energy configuration. 
\section*{Acknowledgements}
We thank Kohei Kamada and Igor Shovkovy for very useful discussions on analytical approaches to chiral MHD turbulence. 
This work was supported by the Deutsche Forschungsgemeinschaft through the Collaborative Research Centre SFB
676 ”Particles, Strings and the Early Universe”.

.
\bibliography{apssamp}

\end{document}